\documentclass[dvips]{article}
\usepackage{icrctc07}

\title{Search for pulsed multi-TeV gamma rays from 
the Crab pulsar using the Tibet-III air shower array
}
\shorttitle{Search for pulsed multi-TeV gamma rays}
\authors{
The Tibet AS$\gamma$ Collaboration\\
M.~Amenomori$^{1}$, X.~J.~Bi$^{2}$, D.~Chen$^{3}$, S.~W.~Cui$^{4}$,
Danzengluobu$^{5}$, L.~K.~Ding$^{2}$, X.~H.~Ding$^{5}$, C.~Fan$^{6}$,
C.~F.~Feng$^{6}$, Zhaoyang Feng$^{2}$, Z.~Y.~Feng$^{7}$,
X.~Y.~Gao$^{8}$, Q.~X.~Geng$^{8}$, H.~W.~Guo$^{5}$, H.~H.~He$^{2}$,
M.~He$^{6}$, K.~Hibino$^{9}$, N.~Hotta$^{10}$, Haibing~Hu$^{5}$,
H.~B.~Hu$^{2}$, J.~Huang$^{11}$, Q.~Huang$^{7}$, H.~Y.~Jia$^{7}$,
F.~Kajino$^{12}$, K.~Kasahara$^{13}$, Y.~Katayose$^{3}$,
C.~Kato$^{14}$, K.~Kawata$^{11}$, Labaciren$^{5}$, G.~M.~Le$^{15}$,
A.~F.~Li$^{6}$, J.~Y.~Li$^{6}$, Y.-Q.~Lou$^{16}$, H.~Lu$^{2}$,
S.~L.~Lu$^{2}$, X.~R.~Meng$^{5}$, K.~Mizutani$^{13,17}$, J.~Mu$^{8}$,
K.~Munakata$^{14}$, A.~Nagai$^{18}$, H.~Nanjo$^{1}$,
M.~Nishizawa$^{19}$, M.~Ohnishi$^{11}$, I.~Ohta$^{20}$,
H. Onuma$^{17}$, T.~Ouchi$^{9}$, S.~Ozawa$^{11}$, J.~R.~Ren$^{2}$,
T.~Saito$^{21}$, T.~Y.~Saito$^{22}$, M.~Sakata$^{12}$,
T.~K.~Sako$^{11}$, M.~Shibata$^{3}$, A.~Shiomi$^{9,11}$,
T.~Shirai$^{9}$, H.~Sugimoto$^{23}$, M.~Takita$^{11}$,
Y.~H.~Tan$^{2}$, N.~Tateyama$^{9}$, S.~Torii$^{13}$,
H.~Tsuchiya$^{24}$, S.~Udo$^{11}$, B.~Wang$^{8}$, H.~Wang$^{2}$,
X.~Wang$^{11}$, Y.~Wang$^{2}$, Y.~G.~Wang$^{6}$, H.~R.~Wu$^{2}$,
L.~Xue$^{6}$, Y.~Yamamoto$^{12}$, C.~T.~Yan$^{11}$, X.~C.~Yang$^{8}$,
S.~Yasue$^{25}$, Z.~H.~Ye$^{15}$, G.~C.~Yu$^{7}$, A.~F.~Yuan$^{5}$,
T.~Yuda$^{9}$, H.~M.~Zhang$^{2}$, J.~L.~Zhang$^{2}$,
N.~J.~Zhang$^{6}$, X.~Y.~Zhang$^{6}$, Y.~Zhang$^{2}$, Yi~Zhang$^{2}$,
Zhaxisangzhu$^{5}$ and X.~X.~Zhou$^{7}$
}

\shortauthors{M.~Amenomori {\it et al.}}

\afiliations{\noindent
$^{1}$Department of Physics, Hirosaki University, Hirosaki 036-8561, Japan.
$^{2}$Key Laboratory of Particle Astrophysics, Institute of High Energy Physics, Chinese Academy of Sciences, Beijing 100049, China.
$^{3}$Faculty of Engineering, Yokohama National University, Yokohama 240-8501, Japan.
$^{4}$Department of Physics, Hebei Normal University, Shijiazhuang 050016, China.
$^{5}$Department of Mathematics and Physics, Tibet University, Lhasa 850000, China.
$^{6}$Department of Physics, Shandong University, Jinan 250100, China.
$^{7}$Institute of Modern Physics, SouthWest Jiaotong University, Chengdu 610031, China.
$^{8}$Department of Physics, Yunnan University, Kunming 650091, China.
$^{9}$Faculty of Engineering, Kanagawa University, Yokohama 221-8686, Japan.
$^{10}$Faculty of Education, Utsunomiya University, Utsunomiya 321-8505, Japan.
$^{11}$Institute for Cosmic Ray Research, University of Tokyo, Kashiwa 277-8582, Japan.
$^{12}$Department of Physics, Konan University, Kobe 658-8501, Japan.
$^{13}$Research Institute for Science and Engineering, Waseda University, Tokyo 169-8555, Japan.
$^{14}$Department of Physics, Shinshu University, Matsumoto 390-8621, Japan.
$^{15}$Center of Space Science and Application Research, Chinese Academy of Sciences, Beijing 100080, China.
$^{16}$Physics Department and Tsinghua Center for Astrophysics, Tsinghua University, Beijing 100084, China.
$^{17}$Department of Physics, Saitama University, Saitama 338-8570, Japan.
$^{18}$Advanced Media Network Center, Utsunomiya University, Utsunomiya 321-8585, Japan.
$^{19}$National Institute of Informatics, Tokyo 101-8430, Japan.
$^{20}$Tochigi Study Center, University of the Air, Utsunomiya 321-0943, Japan.
$^{21}$Tokyo Metropolitan College of Industrial Technology, Tokyo 116-8523, Japan.
$^{22}$Max-Planck-Institut f\"ur Physik, M\"unchen D-80805, Deutschland.
$^{23}$Shonan Institute of Technology, Fujisawa 251-8511, Japan.
$^{24}$RIKEN, Wako 351-0198, Japan.
$^{25}$School of General Education, Shinshu University, Matsumoto 390-8621, Japan.
}

\abstract{
We searched for pulsed gamma-ray emissions 
from the Crab pulsar using data of the Tibet-III air shower array 
from November 1999 through November 2005. 
No evidence for the pulsed emissions was found in our analysis.
Upper limits at different energies were calculated for 
a $3 \sigma$ confidence level in the energy range of multi-TeV to 
several hundred TeV.
}

\begin{document}
\maketitle

\section{Introduction}

The Crab Nebula is one of the most studied objects and is the most energetic source 
at GeV - TeV energies.
The energy source of that activity is known to be a pulsar in the nebula.
The rotation period of the Crab pulsar is 33 ms, as inferred from radio, light, and X-ray observations.
Pulsed emission with that rotation period has been detected by EGRET on board 
CGRO \cite{Fierro} at GeV energies,
although several observers have reported no evidence of pulsed emissions of 
greater than 10 GeV energy \cite{CELESTE,HEGRA,Whipple,HESS,MAGIC}.

The emission models of high-energy pulsed gamma rays are mostly the polar cap 
\cite{PolarCap} and the outer gap \cite{OuterGap} models.
Their models predict a sharp cutoff of the energy spectrum because of the 
limitation of particle acceleration. 
The expected cutoff energy depends on many parameters of each model. 
Those parameters would be determined by observations.
Herein, we present the results of a search for pulsed gamma rays from the Crab pulsar 
at energies of 2 TeV to 200 TeV using a Tibet-III air shower array.

\section{Experiment}

The Tibet-III air shower array used in this experiment was constructed in 1999
 at Yangbajing (4300 m a.s.l.) in Tibet.
The array, corresponding to the inner part of the full-scale Tibet-III air shower array,
 consists of 533 scintillation counters covering 22 050 m$^2$ \cite{Amenomori2003,Science2006}.
The mode energy of detected events is about 3 TeV for proton-induced showers 
and the angular resolution is 0.9$^\circ$. The systematic error of the energy determination
of primary particles and systematic pointing error of the array were well calibrated by comparing the observed displacement of the moon shadow 
because of the geomagnetic field with the Monte Carlo simulation, as described
in a previous paper \cite{Kawata}.

We observed steady excess events from the Crab Nebula during 
November 1999 through November 2005. 
These events were selected by imposing the following conditions:
1) each shower must fire four or more counters recording 1.25 or more particles; 
2) all fired counters or eight of nine fired counters which recorded the
highest particle density must be inside the fiducial area; and 
3) the zenith angle of the arrival direction must be less than $ 40^{\circ}$.
After these selections, the events were examined for further analyses.

\section{Data Analysis}

The data analyzed here were chosen for events coming from 
a window around the direction of the Crab pulsar.
The search window radius is expressed as $6.9 / (\sum \rho_{FT})^{1/2}$ (degree),
where $\sum \rho_{FT}$ is the sum of the number of particles m$^2$ for each 
scintillation counter with a fast-timing (FT) PMT.
The function was optimized to maximize the $S / \sqrt{N}$ ratio using MC simulation
 \cite{Amenomori2003}.
 
The arrival time of each event is recorded using a quartz clock synchronized with GPS, 
which has a precision of 1 $\mu$s.
For the timing analysis, all arrival times are converted to the solar system barycenter frame 
using the JPL DE200 ephemeris \cite{Standish1982}.

The Crab pulsar ephemeris is calculated using the Jordrell Bank Crab Pulsar Monthly 
Ephemeris \cite{JBC0,JBC1}.
The corrected arrival time of each event is calculated to a rotated phase of the Crab 
pulsar, which takes account of the derivative $\dot{P}$ of the period $P$ month by month.

\section{Results}

\begin{figure}
\begin{center}
\includegraphics [width=0.48\textwidth]{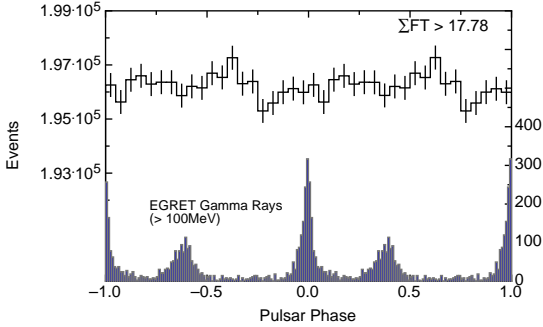}
\end{center}
\caption{
Distribution of the event phase of the Crab pulsar.
Phase 0 is defined using the timing solution derived from the main pulse 
of the radio observations.
The upper plot shows our result for $\sum \rho_{FT} > 17.78$.
The lower plot shows the $\gamma$-ray phase histogram above 100 MeV,
as measured using EGRET \cite{Fierro}.
}
\label{fig1}
\end{figure}

\begin{figure}
\begin{center}
\includegraphics [width=0.48\textwidth]{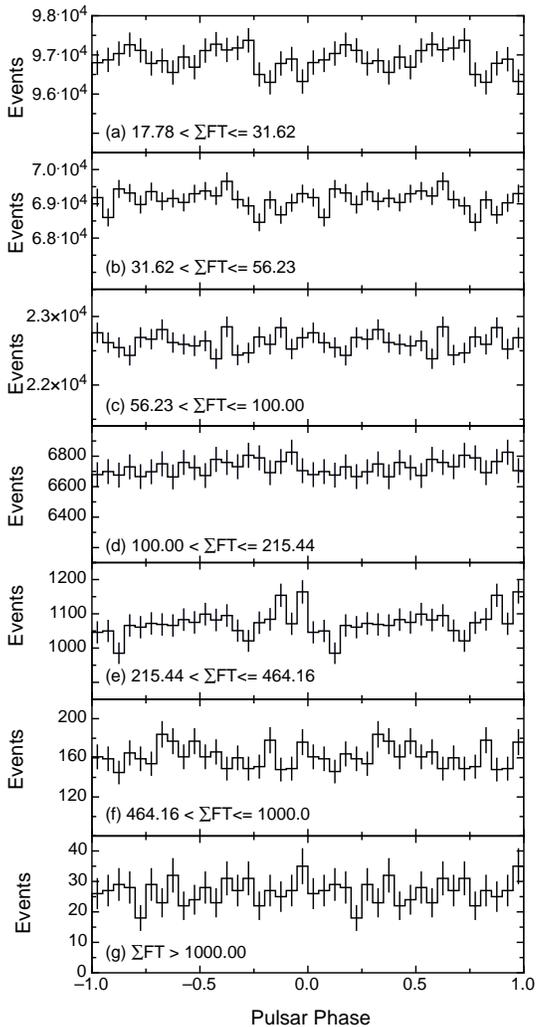}
\end{center}
\caption{
Distributions of the event phase of the Crab pulsar.
Each plot shows a histogram for every $\sum \rho_{FT} $ range,
which is the equivalent of the energy region, as shown in Table \ref{result}.
}
\label{fig2}
\end{figure}

Figure \ref{fig1} shows the distribution of events for each phase in two rotational 
periods of the Crab pulsar. 
The distribution is compatible with a flat distribution ($\chi^2/d.o.f. = 0.95$). 
That is, no significantly pulsed signal was found in observations with mode energy of $\sim$ 3 TeV.
The phase analysis is performed on seven 
intervals of $\sum \rho_{FT}$, as shown in Fig. \ref{fig2} to examine the energy dependence.
Table \ref{result} shows the statistical results of the applied $Z^2_2$ test \cite{z-test} 
and $H$ test \cite{h-test}, as well as the $\chi^2$ test.

\begin{table*}[hbtdp]
\caption{Results of statistical tests for pulsed emission.
$\chi^2$-, $Z^2_2$- and $H$-test (probabilities) are calculated for a flat phase distribution.}
\begin{center}
\begin{tabular}{ccccc}
\hline 
$\sum \rho_{FT} $ & Energy (TeV) & $\chi^2/d.o.f.$ & $Z^2_2$ & $H$ \\
\hline\hline
17.78 -- 31.62 & 2.1 & 0.97 (0.49)   & 9.62 (0.047) &  9.62 (0.021)\\
31.62 -- 56.23 & 3.6 & 1.21 (0.24) & 7.64 (0.11) &  7.64 (0.047)\\
56.23 -- 100.00 & 5.7 & 0.81(0.70)  & 2.54 (0.64) & 4.49  (0.17) \\
100.00 -- 215.44 & 9.3 & 0.35 (0.96) & 2.30 (0.68) & 6.14  (0.086) \\
215.44 -- 464.16 & 20.4 & 1.41 (0.11) &  9.68 (0.046) &  14.56 (0.0030) \\
464.16 -- 1000.0 & 51.7 & 0.80 (0.71) & 3.67 (0.45) & 6.09  (0.088) \\
$>$ 1000.0 & 122.7 & 0.60 (0.91) & 1.11 (0.89) & 4.48  (0.17) \\
\hline
$>$ 17.78  &  $>$ 2.1 & 0.95 (0.52) & 8.41 (0.078) & 8.87  (0.029) \\
\hline
\end{tabular}
\end{center}
\label{result}
\end{table*}%

\begin{figure}
\begin{center}
\includegraphics [width=0.48\textwidth]{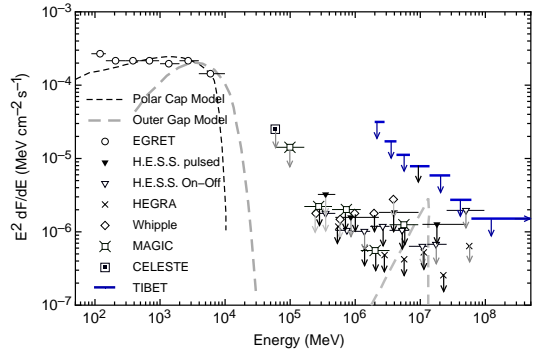}
\end{center}
\caption{
Upper limits on the pulsed gamma ray flux from the Crab pulsar.
}
\label{fig3}
\end{figure}

Almost all statistical test results show that the phase distributions are uniform
within a 3 $\sigma$ significance level.
We have estimated the $3 \sigma$ flux upper limit of the pulsed emission from the 
Crab pulsar using the $H$ test \cite{h-test} as

\begin{eqnarray*}
x_{3\sigma} & = & (1.5+10.7\delta)(0.174H)^{0.17+0.14\delta}\\
 & & \times  \exp \{(0.08+0.14\delta) \\
 & & \times (\log_{10}(0.174H))^2\},
\end{eqnarray*}

where $\delta$ is the duty cycle of the pulse component, 
assuming the $\delta$ for the Crab pulsar is 21\%.
Exposure from the Crab pulsar for the Tibet-III experiment
is estimated using MC simulation.
The upper limit is compared to previous results inferred from results of other experiments, 
as shown in Fig. \ref{fig3}.

\section{Conclusions}

During the period from November 1999 to November 2005,
we searched for pulsed gamma-ray emissions synchronized 
with the rotational period provided from the radio observation of the Crab pulsar. 
No evidence for the pulsed emission was obtained through our analyses.
The upper limits at different energies were calculated for 
a $3 \sigma$ confidence level.
These results are inconclusive in relation to the polar cap and outer gap model.
We will report additional detailed analyses and discussion in the near future.

\section{Acknowledgements}

The collaborative experiment of the Tibet Air Shower
Arrays has been performed under the auspices of the
Ministry of Science and Technology of China and the
Ministry of Foreign Affairs of Japan. This work is 
supported in part by Grants-in-Aid for Scientific Research
on Priority Areas (712) (MEXT) and by Scientific Research (JSPS) in Japan,
and by the Committee of the Natural Science
Foundation and by the Chinese Academy of
Sciences in China.

\bibliography{icrc0844}
\bibliographystyle{unsrt}

\end{document}